# Text Compression And Superfast Searching

Udayan Khurana[1]

Anirudh Koul[2]

[1,2] Thapar Institute Of Engineering and Technology, Patiala, Punjab, India-147004

## ABSTRACT

In this paper, a new compression scheme for text is presented. The same is efficient in giving high compression ratios and enables super fast searching within the compressed text. Typical compression ratios of 70-80% and reducing the search time by 80-85% are the features of this paper. Till now, a trade-off between high ratios and searchability within compressed text has been seen. In this paper, we show that greater the compression, faster the search. This finds applicability in so many places where data as natural language text is present.

## 1. INTRODUCTION

In this age of information and in the era of distributed computing, one thing that is growing at an exponential rate is information itself. As more and more nodes become a part of the Internet community, data sharing grows as never before. Increasing Office Automation, Digital Libraries, Meta-data storage and explosion in the amount of communication suggest that it is indeed a wise option to employ the best possible compression utilities. For instance, as much data is going to produced in the next two years as the entire human civilization has had till now. And 97% of that is going to be computerized. Search engines that store every page on net, in today's date store approximately 20-30TeraBytes of data! Information storage and exchange suggest that natural language text is omnipresent. It is thus worthwhile to consider schemes that though not generic are specifically applicable to text for superior results.

This paper deals with a new compression technique developed for text data. It uses the fact that text of a language would essentially contain certain set of binary strings more than other, i.e. it takes advantage of the fact that the data is text. The results produced for text are superior to those produced by other generic or specialized techniques in popular use. Where have many other compression techniques been non-usable for archiving and data sharing is that retrieving any search results becomes quite an uphill task because of Decompression first and then searching. This technique of ours is quite suitable in performing the search tasks even within the compressed file. The encryption used makes the average search even faster. In this text, we also demonstrate the searching feature by giving an algorithm for the same.
Related Work

Compression algorithms have a long list. I this paper we shall consider only the lossless compression schemes and not the lossy ones because of relevance to our subject. The starting work was done by Huffman [1], where pieces of data (segments or bytes) in a file are replaced by a code of shorter length. There exist the LZ family compression algorithms [5], [6], variants of which are popularly used today. Arithmetic coding is an useful technique that gives good compression ratios [4]. Other compression schemes include index compression. Manber [8] introduced a technique where one could pair the frequently occurring adjacent characters. Though the compression produced is low (typically

---

[1] The Algorithm presented in the paper is an invention of the First author
[2] Second Author has contributed in Code Implementation and Results Collection

30%) but it provides a flexible search algorithm within the compressed text. Recently, there has also been a lot of research on finding efficient searching algorithms within files compressed by LZ family [2], [3]. Many searching algorithms use indexing for faster results.
None of the algorithms provides desirable results. The searching algorithms that operate on LZ compressed files are at best O(min(u,mn) + n[w/r] + R) worst case and O(min(u,mlogn) + n[w/r] + R) average case. Zivizni et al. [7] state a possible problem with searching Huffman Compressed files that a string may appear after the search even when it did not exist in the original text.
We provide a scheme to compress text where good ratios of compression (typically 70%) are achieved and searching time decreases by a whopping 80 %.
It can be summed up in the following table:

| Compression Technique | Compression Results | Search |
|---|---|---|
| Huffman | Bad-OK | Bad-OK |
| Index Compressions | Bad | OK |
| LZ family | Good | Bad |
| Ours | Good | Good |

Table 1. A relative categorization of various techniques with respect to compression and search results

## 2. COMPRESSION

(All the references taken in the following text are with respect to the English Language and should easily be extensible to most of the other languages as well)

### 2.1 Basic idea behind this technique

Typically, a natural language text consists of words and punctuations forming up sentences. Most of the words belong to a standard list called the dictionary for that language. (Note: The term dictionary refers to the dictionary of general use and not the one referred to by many compression techniques). We shall call this the referencing list. Since most of the words in a text are dictionary based, why not use a standard reference list and so index (or refer) to each present word by a reference number. Upon calculations and estimates, we found that it would be best to use a two-byte reference number. (Experimental results leading to the above conclusion are being omitted because of limitation of space. However, one may notice that there are about 40 thousand words in an English Language Dictionary and as one reeds on further about 15 thousand indexes should be needed to comprehensively take care of every kind of incoming text.) So, possibly $2^{16}$= 65,536 combinations that may be referred to using an index of length 2-Bytes. (Notice that the number lesser than this would be 15 bits that may refer to only 32,768 entities). The length of this index is fixed through out..

### 2.2 COMPRESSION SCHEME

Based upon statistics of usage, average word length in English Language is **4.3 c**haracters in length (Statistics omitted due to limitation of space). There is a space between two words, that makes the average word length = **5.3.** Using an **extended-ASCII** coding, number of bytes that are consumed to write an average word = **5.3** Bytes. If a two-byte referencing number is used , referring to each word in the text by a 2-Byte integer, each word now occupies 2-Byte space. So, in a very naïve calculation, one may say, compression = (5.3 – 2)/5.3= **62 %** compression straight away. The above arithmetic shows the basic theme of this compression scheme, which utilizes the fact that text from a language may belong to (for the most part) a limited set of entities (limited does not mean a small number). ( For illustrations refer to the appendix)

## 2.3 WHAT IF THE WORD LIES OUT OF THE REFERENCING LIST

A text would often contain words that cannot be the part of a standard reference. Those may be: Names of people, places (specially those that are not very common), a new term that has come into existence recently or an irregularity, (may be a typing error).To accommodate such occurrences, we have devised the referencing scheme*, so as to take care that no unknown text is stored with a negative compression ratio.
** Refer to the Complete Indexing Scheme.*

### 2.3.1 ACCOMMODATION WITHIN A REFERENCE INTEGER

In this subsection, we consider how is it possible to code for a new word that does not lie in our referencing list. We aim to use index numbers from within our Index range and use each 2-Byte index to adjust two bytes of a new word, thereby ending up in a no profit no loss deal in the worst case. Number of characters in roman script ( 'a'-'z', A'-'Z'), Numerals (0-9), 1 Nothing character * = 52+ 10+1 = 63. Square of 63= 63*63= 3969 combinations. We would accommodate 2 characters of a new word in each index that represents an ordered pair. Hence we shall refer to an ordered pair of characters by **3969** combinations. So, we may have, 56,500 to 60,469 representing that a new word these is the starting of a new word and mentions its **first two characters**. Also 61000-64696 suggests the continuation of a new word from the last index and mentions its **next two characters**. It is significant to represent the starting and continuation by different index sets (classes). This is useful, when two consecutive words are not in the reference list L.

### 2.3.2 PURELY NUMERIC WORDS NOT IN REFERENCE LIST

It is worth using a different scheme to encode purely numerical words that lie outside the reference scheme. This is necessarily because of the lesser domain of a decimal alphabet and we show that the local compression is higher when a scheme separate from alphanumeric words is used. As we did for alpha-numeric characters words, we form triplets of purely numeric words, We use, 11*11*10 = 1210 combinations for three digits in a single index. So, in all, 1210 + 1210 = 2420 combinations for starting and continuation of a new word shall be used. For instance, we have used index numbers 53500-56000 for this purpose. (Refer to the Appendix)

### 2.3.3 EFFECT ON COMPRESSION

It is important to have knowledge of the local compression when a new word is being encoded. We prove that even the worst case is not going to result in local expansion. In particular, where n is the new word length: For alpha numeric words :
Fractional age compression = 1/n in case of even characters
= 0 in case of odd characters.
Hence, for new words, there is no negative compression.
For purely numeric words, Fractional compression: (For Proof, please refer to the Appendix)
= (n+3)/(3n+3) in case n=3k
= 1/3 in case n=3k+2
= (n-1)/3n+3 in case n=3k+1
where n is the new word length and k is an integer >=0.

## 3. THE ALGORITHM

The algorithm is divided into four phases or Parses. Parse I takes as input the uncompressed text and each parse produces an output that serves as input to the next parse, un till the final one. Parse 3 is not implemented as yet and parse four is an optional one. It though improves compression but retards the searching speed.

Parse I
1.Scan the whole text and break down words separated by either space or special symbols or a combination of both. For each word and each separator, follow steps 2 to 4.
2.Search whether the word lies in the reference list or not. If yes place the reference number/index in the compressed file.
3.For each space between two words, move on to the next word.
4.For each combination of spaces and special symbols, place the particular code for each combination.

Parse II
1. Scan the compressed file of phase I to find combinations of two or more words as are present in the combination list.
2. Replace each such combination by the appropriate index number.

Parse III
1. Find for each possible sentence structure amongst those listed in the sentence list.
2. Replace the appropriate index number with the required parameters that follow in the consecutive index numbers.
3. List the frequency of each occurring index in a separate data structure.

Parse IV
1. Find patterns of repetitive occurrence of sequence of indexes.
2. Define an alias for each (within the specified index range) against the first occurrence of each.
3. Replace the appropriate sequence of index numbers with the alias.

## 4. MOVING ABOVE THE 60%

a)Following connotations are often noticed while the language is in use,
we are      in the      is of      is a      I am      because of
and so many others, using which as single entities, gives rise to higher compression ratios. It is obvious that, smaller length words are lesser compressed. This way (such combinations), increase the average length of a compressible entity, and hence increase the compression ratio. We shall call these as combined words or composite words and hereon, C denotes their complete set. In specialized information systems, the data is specialized, so there occur more of larger blocks of data that can be treated as a single entity. This typically doubles up the local compression ratio. (For an illustration, refer to the appendix)

b) Placing the most frequently occurring words in a cluster of 256 indexes such that every time a word among those has been referred, first eight bits out of the total 16-bit index are bound to be the same. This generates additional redundancy, which is made use of in the Parse VI of compression algorithm to produce higher ratios. ( Refer to appendix for an illustration)

## 5. PROCEDURE FOR SEARCHING A PHRASE IN THE COMPRESSED FILE

A piece of text is easily searchable within a compressed file (without the need of decompression):

i. Each word essentially has a corresponding integer that can be mapped from the Reference List L.
ii. Since there are no cross-linkages within a compressed file, by searching the corresponding index number, it sufficient to test the presence of a match.

### 5.1 ALGORITHM

Assumption 1: Phase I, II compressed file
Assumption 2: phrase is a single word (combinations should work similarly)
The symbol (quotes for clarity) ‘ ζ’ means, “ is a subset of”

Let W be the word to be searched. COMP is the compressed(encrypted) file. L is the list of (single) words references.
C is the list of composite words references. $C_I$ is the set of individual words that compose elements of C. Also, $C_I$ ζ L
$C_W$ is the set of elements in C that contain W. NL is the set indices referring to a new word i.e. not in L ( and hence not in $C_I$). R is an ordered sequence of one or more indexes that describe a word not belonging to L. $R_I$ Individual index numbers, s.t. all $R_I$ ε NL . Also, for n>0, ordered set { $R_{I1}$,$R_{I2}$ .... ,$R_{In}$} ζ R

ALGORITHM SEARCH

---

Step 1) If the W ε $C_I$ ,
Then k=getindex($C_W$)
Search for k in COMP.
Step 2) If W ε L
Then k=getindex(W)
Search for k in COMP
Step 3) If !((W ε L) && (W ε $C_I$ )) // From Step 1 & Step 2
Then,
Search for all R in COMP,
Convert all R sequences corresponding to each word to their original meaning and compare W as ordinary. Repeat for all values R found.

---

## 5.2 HOW DOES IT MAKE SEARCH FASTER

What follows is an estimate of the searching time requirements. To search a word (or any pattern of bytes) in the compressed text, one needs to compare all the indexes present in the compressed text after translation. The amount of time required for translation is essentially fixed for every translation and is small enough to be significant. Searching the index in the file would be require comparisons equal the number of the total indexes in the compressed file. Searching in the uncompressed file would take comparisons as many as at the least the total number of characters in the file.
Let the total number of comparisons in the uncompressed file be 5.3x. Then the number of compressions in the compressed file would be: x. (Since for every 5.3 characters in the original file, there is one index number in the compressed file). Therefore, the percentage saving in time would be = (5.3x-x)/5.3x = 81 %. We call this number as the Search Boost Factor (SBF). For practical estimation, in a fractional compression z: SBF= 50 * (1 +z). (For proof, refer to Appendix)

## 6. RESULTS

## 6.1 COMPRESSION

| File | Original Size (KB) | Compr-essed Size(B) | Compress-ion | | US Patent Number | Original Size(B) | Compressed Size (B) | Compr-ession |
|---|---|---|---|---|---|---|---|---|
| Text 1 | 14.8 | 3846 | 74.62% | | 6,227,334 | 1246 | 304 | 75.60% |
| Text 2 | 1302.12 | 346189 | 74.03% | | 6,227,354 | 1538 | 346 | 77.50% |
| Text 3 | 573.212 | 131682 | 77.60% | | 6,227,352 | 584 | 137 | 76.50% |
| Text 4 | 462.74 | 121139 | 74.43% | | 6,237,576 | 230 | 55 | 76.10% |
| Text 5 | 310.79 | 81557 | 74.37% | | 6,227,332 | 276 | 64 | 76.80% |

Table 2. Compression Results For Some E-books Table 3. Compression Results For Small Files

Text 1 :GPL  Text 2: The Three Musketeers Text 3 : Anne Of Green Gables Text 4 : Heroes of Telegraph Text 5 : History of Telephone

### 6.2 SEARCHING

| SIZE (Bytes) | | | SEARCH TIME (ms) | | |
|---|---|---|---|---|---|
| Original | Compressed | %Compression | Original | Compressed | % Saving |
| 5014247 | 1420802 | 72 % | 447 | 73.4 | 84% |
| 2507126 | 6998952 | 72% | 207.7 | 37 | 82% |
| 1253565 | 330117 | 74% | 105 | 17.7 | 83% |

Table 4. Searching Results Compared to linear search in the Uncompressed Text
Note : The results were obtained on a 1.5 GHz Pentium M IBM R51 Thinkpad, (OS: Windows XP)

## 7. APPLICATION

The sphere of application is indeed very large. Data systems with specialized dictionaries are bound to have more specific text and several occurrences are more frequent that should give rise to even better compression ratios. This compression scheme can be a feature of word processors, archiving systems, email storage and every other place where natural language text is involved. It may also prove useful for reducing bandwidth for static and mobile networks, as reducing the message size lowers the network requirements. Important features of this technique would be no overhead occupation of space as is present in other compression techniques to build a dictionary and of course the consistency of compression (even files without considerable redundancy are well compressed. For details refer to appendix).

## 8. CONCLUSION AND SCOPE FOR FURTHER RESEARCH

In this paper, a new compression technique that uses referencing through two-byte numbers (indices) for the purpose of encoding has been presented. The technique is efficient in providing high compression ratios and faster search through the text. It leaves a good scope for further research for actually incorporating phase 3 of the given algorithm. The same should need extensive study of general sentence formats and scope for maximum compression. Another area of research would be to modify the compression scheme so that searching is even faster. Incorporating indexing so as to achieve the same is yet another challenge. Any reviews and research proposals are welcome.

## 10. APPENDIX

### APPENDIX I - EXAMPLES, PROOFS AND EXPLANATION

**1) Example to illustrate the basic nature of Compression**
For e.g. the sentence ***"The quick brown fox jumped over the lazy dog"***
Would be compressed up to 60% as per this scheme, using 9 indexes to represent 9 words that appear in this sentence. Where as the other typical techniques do not compress it at all because it does not contain any repetition of bytes or segments. So, an important point to be noted here is that data need not be redundant in nature to be compressible. Hence small texts such as emails and SMS maybe compressed quite effectively which is generally not the case because of large overheads denying a profitable use of some compression.

**2) THE NOTHING CHARACTER**
The nothing character is one that denotes that the word has ended and in this place there is no character. This is brought in practice when a new word is of odd number of characters. We know, every index, in this case, stores an ordered pair of characters. Hence, the last character is stored as a combination (ordered) of that character and a nothing character code.

**3) Effect on Compression when a new word is met (Alphanumeric):**
Effect on compression
It is important to have knowledge of the local compression when a new word is being encoded. We prove that even the worst case is not going to result in local expansion. In particular, where n is the new word length:
Fractional age compression = 1/n in case of even characters
= 0 in case of odd characters.

Proof: There are two possible cases: n being even and odd, depending on which the local compression varies.
For even n,
Bytes required in the original storage = n
Number of indexes required, = n/2.
No. of bytes in compressed storage= 2 * n/2 = n
Hence, fractional compression = (n+1 – n)/n = 1/n (1)
For odd n,
Bytes required in the original storage = n
Number of indexes required, = (n+1)/2.
No. of bytes in compressed storage= 2 * (n+1)/2 = n+1
Hence, fractional compression = (n+1 – n+1)/n+1 = 0 (2)

*Hence, from (1) and (2), we infer that for new words, there is no negative compression.*

**4) Effect on Compression when a new word is met (Purely Numeric):**

Fractional compression = (n+3)/(3n+3) in case n=3k
= 1/3 in case n=3k+2

= (n-1)/3n+3 in case n=3k+1

where n is the new word length and k is I >=0.

Proof : For n=3k

Bytes required in the original storage = n+1

Number of indexes required, = k=n/3.

No. of bytes in compressed storage= 2 * n/3 = 2n/3

Hence, fractional compression = (n+1 – 2n/3)/n+1 = (n+3)/(3n+3 ) (3)

For n=3k+2,

Bytes required in the original storage = n+1

Number of indexes required, = k+1=(n+1)/3.

No. of bytes in compressed storage= 2 * (n+1)/3

Hence, fractional compression = (n+1 – 2(n+1)/3)/(n+1) = 1/3 (4)

For n=3k+1,

Bytes required in the original storage = n+1

Number of indexes required, = k+1=(n+2)/3.

No. of bytes in compressed storage= 2 * (n+2)/3 = (2n+4)/3

Hence, fractional compression = (n+1 – (2n+4)/3)/n+1 = (n-1)/(3n+1) (5)

Hence, from (3), (4) and (5), we infer that for new numeric words, there is no negative compression.

**5) COMMON COMBINATIONS**

In the very frequent combination "***in the*** ",

We have, the combined length of the compressible entity = 2+1+3+1= 7.

Hence compression % = (7-2)/7 *100= **71 %**.

Instead of the earlier, (7-4)/7*100= **42%**.

Hence, the net compression also moves up.

**6) FREQUENT WORDS AND ADDED REDUNDANCY**

We notice that, many words are repeated more often than others :

Words like

*and, the, are, is, a, in, by, have, will, to, of*

and many others

Why not make use of the fact that a limited number of words are used very frequently?

The extended ASCII is an 8-bit encoding. We have been using a 16-bit code for indexing. While writing the compressed file, let us break up the 16-bit index into two 8-bit characters. If we index the common 256 words together, we can produce a redundant data out of the same piece of compressed code!!

Let us keep the indexes of the 256 most common words as: 0 – 255. Consequently, the index numbers of the words at 1st, 2nd and last position are :

1. 00000000 00000001

2. 00000000 00000010

and so on … up till 255.

255. 00000000 11111111

So, every time a *Common Word* is referred to, we get one character out of the two that is a repeated one.

Also, the second character is also repeated, whenever the word itself appears again in the text

*Hence, we have arranged our indexes so as to make out redundancy in the same data.* That's magic**.** This gives an additional feast to the Phase IV.

## 7) General Search Boost Factor ( Proof)

Let a text be fractionally compressed to z.
If A & B are the original and compressed file sizes,
Then,

$$(A-B)/A = z$$
$$\Rightarrow A- B = z.A$$
$$\Rightarrow A-z.A = B$$
$$\Rightarrow B = A. (1-z)$$

Thus B= A.(1-z) is the size of the compressed file. Since the file contains two-byte numbers, the number of indexes is half the size.
i.e., number of indexes = B/2=A.(1-z)/2

While searching in a normal (uncompressed) text, there are at least (best case) as many comparisons as the number of characters.
The number of characters in the file = A (size of the file)
Therefore, the number of comparisons required in the uncompressed file = A
Searching in the compressed file would require comparisons = number of indices.
= A.(1-z)/2

Hence, % age saving in search time = (A – A(1-z)/2 ) / A * 100
Search Boost Factor (SBF) = 50 (1+z) %

Hence, for typical compressions of 70% (z= 0.7)
**SBF = 85 %**

**APPENDIX II**
**OTHER POSSIBLE MEANS OF ENHANCING COMPRESSION**

### 1) SENTENCE GENERALIZATION

2. Sentences with specific structures are often used. Something like a function call with variable parameters may be used to shrink the storage. (not yet implemented) .
e.g. There are *thousands* of *shops*.
There are *places* of *your interest*.
In both the sentences the non –italicized portion is common. So, it may be taken up as a template sentence and the italicized words as arguments.

### 2) UNICODE and COMPRESSION

The adoption of Unicode (I18N) has special effects on this compression strategy, since the same characters would require double the storage (2 bytes for a character) as compared to the earlier one. Multiple dictionaries may be used for different languages within the same file for the purpose if a multilingual text is to be compressed. A similar arithmetic would reveal that compressions of up to 80% and above are easily achievable.

NOTE: ALL THE FOLLOWING RESULTS THIS TEXT ARE BASED UPON *extended-ASCII ENCODING AND NOT UNICODE***.**

## APPENDIX III

### 1) APPLICATION

Such compression schemes have an effect to the mankind. Saving storage space to one-fourth or one-fifth:

- Saves large amounts of inorganic material production like CDs
- Reduces network traffic and hence provides a better infrastructure. (Bandwidth is always limited)
- Reduces cost of large e-mail storage and archives, specially after the provision of 100MB and 1 GB mail spaces.
- This feature can be incorporated in Word Processors, IDEs, Other places where data is involved.

Information Systems which have special and restricted terminology of their own.(an example is given in section 5.4)

### 2) A NOTE ON CONSISTENCY AND PERFORMANCE

It is worth mentioning the consistency in results of this compression technique.

#### NATURE OF COMPRESSION BY REFERENCING

It is quite natural that this technique produces compression consistently. The other techniques that counter redundancy (repetition) to compress some text produce inconsistent results depending on the nature of text. In this technique of ours, the compression is quite independent of that factor. i.e. even a text with no redundancy is as well compressed ( using only first 2 or 3 phases ) as one that is heavily redundant in nature.

#### PERFORMANCE FOR FILES WITH CONSIDERABLE REDUNDANCY

Superior results as compared to other available techniques even in the case of files with heavy repetition ( a delight for other techniques) , comes due to the fact that the Parse IV of our technique counters redundancy ( though not as efficiently as that by others) and always produces a better result. It is thus good enough to say that *this technique performs well at all fronts, whatever might be the nature of data*.

#### GOOD THING ABOUT CONSISTENCY

A consistent compression (at least a lower limit) enables one to estimate the space (average and worst case both) requirement, specially when huge amounts of data has to be stored (typically in case of archiving or email servers).

## APPENDIX IV: THE COMPLETE REFERENCING SCHEME

The list of referencing numbers (0 – 65535) has been broken up into blocks. Each reference index (or number) suggests exactly a single entity i.e. there is no overlap of indices. The blocks of the referencing list are as follows:

| Block Range | Objective | Max number of Values |
|---|---|---|
| 0-255 | Common Words | 255 |
| 256-300 | Repeating word | 45 |
| 301-350 | Repeating space | 50 |
| 351 | New line character | 1 |
| 352 | Tab character | 1 |
| 353-393 | Apostrophe not(Isn’t) | 41 |
| 394-408 | Apostrophe s (he’s) | 15 |
| 409-411 | Apostrophe m (I’m) | 3 |
| 412-426 | Apostrophe ll (we’ll) | 15 |
| 500-1550 | Symbols combination starts | 1050 |
| 1550-2600 | Symbol Combination Repeats | 1050 |
| 2600-2660 | Single letters (52) | 60 |
| 2660-2900 | Other Non Alphanumeric Extended ASCII characters | 240 |
| 3000-52000 | Dictionaries | 49000 |
| 52000-53000 | Common combinations | 1000 |
| 53500-56000 | new number starts | 1210 |
| 55000-56000 | new number repeats | 1210 |
| 56500-61000 | new word starts | 3969 |
| 61000-65031 | new word repeats | 3969 |

Free space has been deliberately kept for extension in the future.